\documentclass[aps,10pt,twocolumn,prl]{revtex4}

\usepackage{graphicx}
\usepackage{verbatim}
\usepackage{color}
\newcommand{\lb}{{<}}
\newcommand{\rb}{{>}}
\begin{document}

\title{Heterogeneity in Classical and Non-Classical Nucleation}

\author{Natali Gulbahce}
\affiliation{Clark University, Department of Physics, Worcester,
MA 01610}
\affiliation{Los Alamos National Laboratory, CCS-3, MS-B256, Los
Alamos, NM 87545}

\author{W.\ Klein}
\affiliation{Boston University, Department of Physics, Boston, MA
02215}
\affiliation{Los Alamos National Laboratory, T-DO, Los Alamos, NM 87545}
\date{\today}

\author{Harvey Gould}
\affiliation{Clark University, Department of Physics, Worcester,
MA 01610}

\date{\today}

\begin{abstract}
We investigate heterogeneous and homogeneous nucleation in nearest-neighbor
and long-range  Ising models for various quench depths. We find that the
system has a true crossover from heterogeneous to homogeneous nucleation for
increasing quench depth only if the interaction is sufficiently long-range.
The survival curves, defined as the fraction of systems that remain in the
metastable state after a given time, have qualitatively different shapes as
a function of quench depth for heterogeneous and homogeneous nucleation when
the interaction is short-range, but have identical shapes within the accuracy
of our data for long-range interactions. 

\end{abstract}

\maketitle

Nucleation, the process by which a metastable state decays, plays an
essential role in a wide variety of systems. The nucleation mechanism
involves the appearance of a critical droplet which overcomes a free energy
barrier and initiates a decay into the stable state. In homogeneous
nucleation the droplet forms due to spontaneous fluctuations. Heterogeneous
nucleation occurs when the droplet forms with the help of a wall, defect, or
an impurity such as an aerosol. Both homogeneous and heterogeneous nucleation
are technologically important, but much greater progress has been made in
understanding homogeneous nucleation~\cite{cahn, langer, uk,kashchiev} than
heterogeneous nucleation~\cite{fletcher, liu, castro, haymet1, haymet2}.

Existing theories of heterogeneous nucleation are phenomenological and assume
that the homogeneous theory can be adapted to the heterogeneous
case~\cite{fletcher, liu, castro, weinberg}. The predictions of these
theories have been compared to experiment, but the comparisons
have been indirect by necessity, and the phenomenological nature of the
theories makes the connection to experiment tenuous.

The existing theories of heterogeneous nucleation assume that the
homogeneous nucleation process is classical, that is, the droplet is assumed
to have a distinct volume and surface and the structure of the droplet
interior is the same as the stable phase~\cite{fletcher, liu,
castro}.  
No theoretical or numerical studies have been
undertaken of the effect of impurities on nucleation near the
pseudo\-spinodal~\cite{heermann, gulbahce}, where in the homogeneous case the
droplets are diffuse with no sharp surface/volume distinction~\cite{uk,
klein2}. This lack is an important omission because many systems such as
metals~\cite{klein1} and polymers~\cite{binder} have a
long-range/near-mean-field nature with important pseudo\-spinodal effects.
We will find that heterogeneous nucleation in near-mean-field systems with
long-range interactions differs in important respects from heterogeneous
nucleation in systems for which homogeneous nucleation is classical.

In this letter we discuss our simulations of the effect of an isolated
impurity and a wall on nucleation. The simulations were done on $d=2$
Ising models at temperature $T=4T_{c}/9$ with an applied magnetic field of
magnitude
$h$. We used the Metropolis algorithm for
about 100 Monte Carlo steps per spin (mcs) to equilibrate the system and then
reversed the field. The magnetization of the stable phase is negative. As
we will discuss, the system generally remains in a metastable state for some
time before the transition to the stable phase. We simulated systems with
interaction range $R=1$ (nearest neighbor) and $R=10$ with linear dimension
$L=100$ (unless otherwise stated), and
$R=20$ with $L=200$. The isolated impurity consists of 5 spins in
the shape of a
$+$ sign, fixed in the direction of the stable phase. The
effect of the impurity on the nucleation rate, the probability of
heterogeneous and homogeneous nucleation as a function of quench depth $h$,
and the structure of the critical droplet was studied.
 
In Table~\ref{tab:prob} we show $f_{\rm impurity}$, the fraction of droplets
that form on the isolated impurity, and $f_{\rm other}$, the fraction of
droplets that form on non-impurity sites, even though an impurity is
present. The number of trials was in the range of 100 to 1000. For both
$R=1$ and
$R=10$ and shallow quenches, that is, close to the $h = 0$ coexistence
curve, the droplet forms on the impurity in all of our trials. Because the
nucleation process is stochastic, we expect that a few droplets would
nucleate on non-impurity sites if we do significantly more trials. As 
$h$ is increased, more droplets form on non-impurity sites. This result was
found for short-range systems using a phase field model~\cite{castro}.

\begin{figure}[ht]
\scalebox{0.7}{\includegraphics{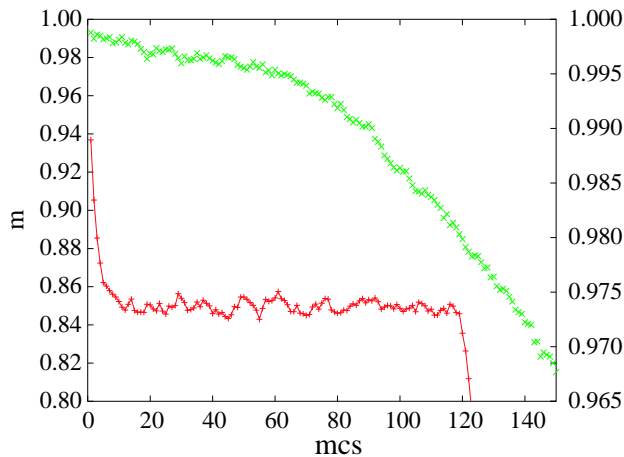}}
\caption{\label{fig:mag} (color online) The magnetization per spin $m$ as a
function of time (mcs) for $R=1$ at
$h=0.40$ ({\color{green}$\times$}) (right axis) and $R=10$ at $h=1.15$
({\color{red}$+$}) (left axis). Note the flat region of $m$ where the $R=10$
system is in metastable equilibrium. There is no metastable equilibrium for
$R=1$ and
$h=0.40$.}
\end{figure}

However, there is a significant qualitative difference for deeper quenches
between the $R=1$ and $R=10$ systems. To see this difference, we plot in
Fig.~\ref{fig:mag} the magnetization per spin $m$ as a function of time for
values of
$h$ where the fraction of droplet formation events on non-impurity spins
becomes significant. For
$R=10$ there is a distinct plateau where $m$ remains roughly constant. This
plateau is associated with the metastable state. However, for
$R=1$ there is no plateau at this quench depth, and thus the system is not
in a metastable state.

\begin{table}[h]
\caption{\label{tab:prob} The fraction of droplet formation events
at spins away from the impurity, $f_{\rm other}$, and on the impurity,
$f_{\rm impurity}$, for $R=1$ and $R=10$ as a function of the quench depth
$h$ for $L=240$. As shown in Fig.~\ref{fig:mag}, the droplet
formation events for deeper
quenches correspond to nucleation only for $R =10$.}
\begin{tabular}{|c|c|c||c|c|c|}
\hline
\multicolumn{3}{|c||}{\bfseries $R=1$} & \multicolumn{3}{c|}{\bfseries
$R=10$}\\\hline\hline
$h$ & ${\rm f_{other}}$ & ${\rm f_{\rm impurity}}$ & $h$ & $\rm f_{other}$
& $\rm f_{\rm impurity}$\\
\hline
0.25 & 0.0 & 1.0 & 1.12 & 0.07 & 0.93\\\hline
0.30 & 0.0 & 1.0 & 1.13 & 0.14 & 0.86\\\hline
0.35 & 0.0 & 1.0 & 1.14 & 0.13 & 0.87\\\hline
0.40 & 0.03 & 0.97 & 1.15 & 0.21 & 0.79\\\hline
0.45 & 0.02 & 0.98 & 1.16 & 0.32 & 0.68\\
\hline
\end{tabular}
\end{table}

In nucleation a critical droplet is a saddle point object, which implies
that it has an equal probability of growing to the stable state or shrinking
back to the metastable state if the system is perturbed at the time of
nucleation~\cite{cahn,langer,uk}. We first ran the simulation until a droplet
formed and the system proceeded to the stable state. The spin configurations
and the current state of the random number generator were saved at various
times. We then chose an intervention time at which the critical droplet might
have appeared, made 20 copies of the system, and restarted the runs with a
different random number seed for each copy~\cite{monette}. If the
intervention time corresponds to the formation of a saddle point object, one
half of the copies will go to the stable phase at approximately the same
time and place as in the original run and one half will return to the
metastable state. If greater than one half return to the metastable state,
the intervention time is too early. If greater than one half proceed to the
stable phase, the intervention time is in the growth phase. Although there
will be significant fluctuations with 20 trials, this number
of trials is sufficient for our purpose.

\begin{table}[ht]
\caption{\label{tab:saddle} The number $n$ of interventions (out of 20) for
which the system goes into the stable phase at approximately the same time
and place as in the original run;
$t$ corresponds to the intervention time when the random number
seed was changed. (The term ``no impurity'' signifies that no impurity was
present in the system.)}
\begin{tabular}{|c|ccccc|}\hline
& \multicolumn{5}{c|}{$t$ (mcs)}\\
\cline{2-6}
& \multicolumn{5}{c|}{$n$}\\ 
\hline\hline
$R=1$ (no impurity) & 140 & 148& 151 & 155 & 160\\
\cline {2-6}
$h=0.6$ & 1& 5 &11& 17& 20\\ \hline
$R=1$ (with impurity)&70 &90 &150&250& 500\\
\cline{2-6}
$h=0.4$ & 10& 5& 15& 10& 20\\\hline
$R=10$ (no impurity) &108 &109 &110 &111 & 112\\ 
\cline {2-6}
$h=1.2$ & 1& 8& 10 & 19& 20\\\hline
$R=10$ (with impurity) & 198& 199& 200& 201& 202\\ 
\cline {2-6}
$h=1.16$ & 3& 5& 15& 19 & 20 \\\hline
\end{tabular}
\end{table}

In Table~\ref{tab:saddle} we give the number of interventions that proceeded
to the stable phase at approximately the same time and place as in the
original run. Note that a saddle point structure is found for the
homogeneous and heterogeneous critical droplets in the deeper quenches for
$R=10$. That is, the fraction of interventions that go to the stable phase is
an increasing function of the time of intervention. Similar behavior was
found for both values of
$R$ for shallow quenches, with and without the presence of an impurity (data
not shown). However, for
$R=1$ in the presence of an impurity and quench depths deep enough so that we
begin to see a non-zero probability of droplet formation at non-impurity
sites, no saddle point structure is found. In this case the fraction of
droplets that proceed to the stable phase at the same time decreases and
then increases as the intervention time increases, in contrast to the
behavior found for
$R=10$ and for $R=1$ in the absence of an impurity. The
lack of a saddle point object for $R=1$ and deep quenches is consistent with
the fact that no plateau was found in Fig.~\ref{fig:mag}. We conclude that
the crossover to a significant probability of homogeneous nucleation 
as the
quench depth increases is found in long-range interaction systems for which
the nucleation is not classical~\cite{uk}. However, in short-range systems the
``crossover'' occurs at quench depths for which the system is no longer
metastable and the decay process cannot be considered to be
nucleation~\cite{binder}. 

Another quantity of interest is the survival curve, which we define as the
fraction of systems that remain in the metastable state after a given
time~\cite{haymet1, haymet2}. We perform this measurement in a slightly
different but equivalent way to that of Heneghan et al.~\cite{haymet1}. We
prepare 1000 systems with the same initial condition, but with different
random number seeds, and then measure the fraction of systems, $\lb s \rb$,
that are in a metastable state for a given value of
$h$ after
$10^{4}$ Monte Carlo steps per spin. 

\begin{figure}[ht]
\scalebox{0.62}{\includegraphics{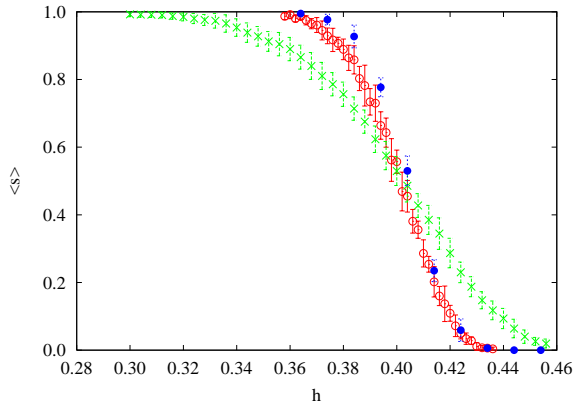}}
\caption{\label{fig:r1} (color online) Survival curves for heterogeneous
nucleation ({\color{red}$\circ$}), homogeneous nucleation (no impurity
present) ({\color{green}$\times$}), and heterogeneous nucleation on a wall
({\color{blue}$\bullet$}) for
$R=1$. The curves for nucleation in the presence of the impurity and on the
wall are shifted to the right by $\Delta h=0.174$ and
$\Delta h=0.255$, respectively, to make it clearer that their shape is
qualitatively different from the homogeneous nucleation curve.} 
\end{figure}

\begin{figure}[ht]
\scalebox{0.62}{\includegraphics{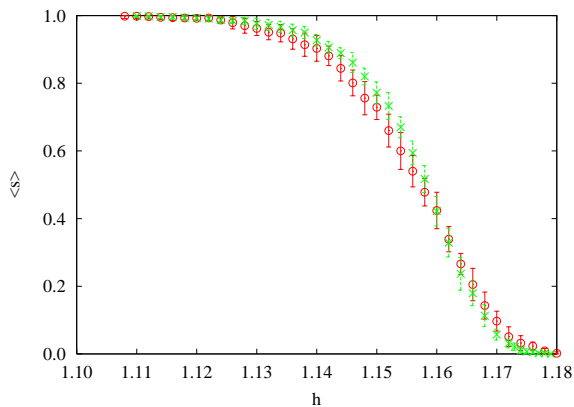}}
\caption{\label{fig:r10} (color online) Survival curves for homogeneous
nucleation ({\color{green}$\times$}) (no impurity)
and heterogeneous nucleation ({\color{red}$\circ$}) for $R=10$. The
latter curve is shifted to the right by $\Delta h = 0.026$ to make it
clearer that the shape of the two curves is almost identical.}
\end{figure}

In Figs.~\ref{fig:r1} and \ref{fig:r10} we plot the survival curves for
$R=1$ and
$R=10$, respectively. Note that for
$R=1$ the shape of the survival curves for heterogeneous nucleation on an
impurity and on a wall is qualitatively different from homogeneous 
nucleation (no impurity present). The wall was
implemented with periodic boundary conditions in one
direction and open boundaries in the other, and a row of fixed spins
at $x=0$. In contrast, for $R=10$, the shape of the survival curves for
homogeneous and heterogeneous nucleation is very similar (see
Fig.~\ref{fig:r10}).

\begin{figure}[ht]
\scalebox{0.68}{\includegraphics{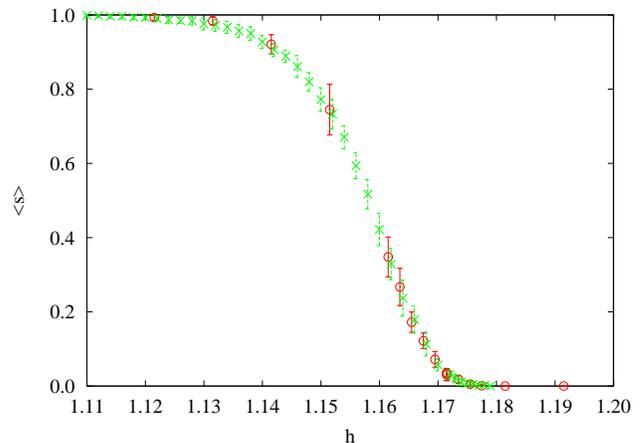}}
\caption{\label{fig:wall} (color online) Survival curves for homogeneous
nucleation ({\color{green}$\times$}) and heterogeneous nucleation
({\color{red}$\circ$}) on a wall for $R=10$. The heterogeneous curve is
shifted to the right by $\Delta h=0.64$.} 
\end{figure}

\begin{figure}[ht]
\scalebox{0.6}{\includegraphics{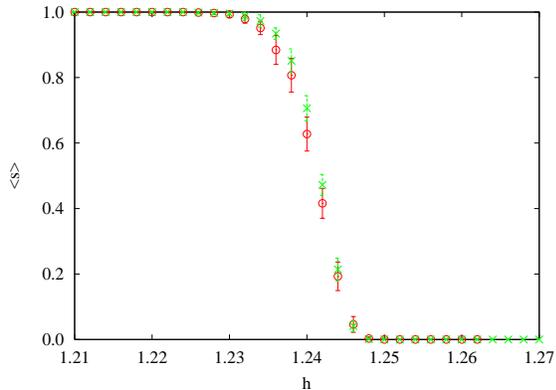}}
\caption{\label{fig:r20} (color online) The survival curves for homogeneous
(no impurity present) ({$\color{green}\times$}) and heterogeneous nucleation
({$\color{red}\circ$}) for $R=20$. The heterogeneous curve is shifted to the
right by
$\Delta h =0.006$ to make it clearer that the two curves have similar
shapes.}
\end{figure}

In Fig.~\ref{fig:wall} the survival curves for $R=10$ for homogeneous
nucleation (no impurity present) and nucleation on a wall are shown. The
curves have been shifted to lay on top of each other and are almost
identical. The results in Figs.~\ref{fig:r10} and \ref{fig:wall} imply
that homogeneous nucleation, nucleation on a wall, and nucleation on
an isolated impurity have survival curves with similar shapes for long-range
interactions. This behavior is reminiscent of the experimental result in
Ref.~\cite{haymet1} where the survival curves for water nucleating on a wall
or on a small crystal of AgI have similar shapes. The survival curves for
$R=20$ are similar to  our results for 
$R=10$ (see Fig.~\ref{fig:r20}). We also did simulations without
fixed spins at $x=0$ and obtained results similar to our other
implementation of a wall (data not shown).

\begin{figure}[ht]
\scalebox{0.6}{\includegraphics{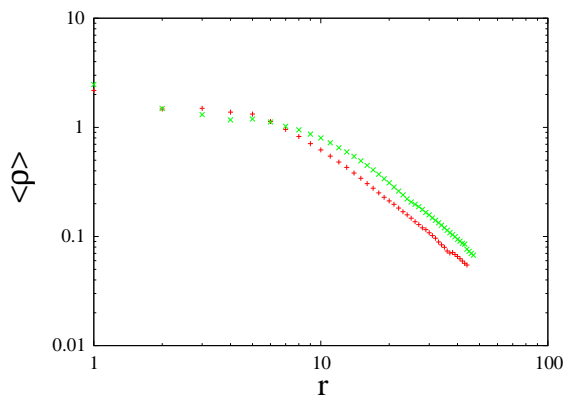}} 
\caption{\label{fig:density} (color online) The average density $\lb \rho
\rb$ of the critical droplet as a function of the radius $r$ measured from
the center of mass for
$R=10$. ({\color{red}$+$} signs indicate heterogeneous
nucleation and
{$\color{green}\times$} signs indicate homogeneous
nucleation.)}
\end{figure}

For
$R=1$ and shallow quenches where the nucleation is classical,
Fletcher~\cite{fletcher} showed theoretically that the radius of the critical
droplet was the same for heterogeneous and homogeneous nucleation. Because
the droplet interior in classical nucleation is the same as the stable
phase, the critical droplets have the same structure aside from the
impurity. However, for
$R=10$, nucleation takes place near the pseudo\-spinodal~\cite{heermann,
gulbahce}, and the critical droplets have a diffuse structure~\cite{uk,
monette}. The question arises as to possible changes in the internal
structure due to the impurity. In Fig.~\ref{fig:density} we plot the density
profiles of the critical droplets that form on and away from the impurity
for $R=10$. Within the accuracy of our data, the structure appears to be the
same.

We now summarize our results and discuss their significance. First,
heterogeneous nucleation droplets appear to be saddle point objects, which
implies that over a significant range of quench depths, simple
modifications of homogeneous nucleation theory is possible. Second, the
common wisdom~\cite{liu, castro} that there is a significant crossover from
heterogeneous to homogeneous nucleation with increasing quench depth appears
to be partially correct. For $R=10$ there is a true crossover because the
system is still in the metastable state, whereas for $R=1$ the ``crossover''
occurs only after the system has been quenched beyond the metastable state. 
The existence of a crossover is important for the theoretical treatment of
nucleation in systems such as metals where there are long-range
interactions~\cite{klein1} and the presence of defects can be important.

We found that the survival curves in long-range systems have the same shape
for both homogeneous and heterogeneous nucleation. This behavior is similar
to the results found experimentally in water~\cite{haymet1}. However, the
survival curve shapes for homogeneous and heterogeneous nucleation are not
the same for $R=1$.

The connection between the survival curve shape and the interaction range
is very difficult to make experimentally because the latter usually cannot
be varied systematically. This connection gives us the first experimentally
accessible marker for when a system exhibits nucleation characteristics
associated with long-range interactions~\cite{uk, monette, klein2}. That is,
if the survival curves have the same shape for homogeneous and
heterogeneous nucleation, near-mean-field effects appear to be important.

We also found that the density profile of the critical droplets for
heterogeneous and homogeneous nucleation is the same within the accuracy of
our data for
$R=10$ and that heterogeneous nucleation is a saddle point process in the
systems we studied and hence can be described by a modified version of
homogeneous nucleation. These results are significant for the calculation
of nucleation rates and will be discussed in a future publication.
 
\begin{acknowledgments}
We acknowledge useful conversations with A.\ D.\ J.\ Haymet,
Greg Johnson, and Frank Alexander. This work was done at Los Alamos National
Laboratory (LA-UR 04-4271). N.\ Gulbahce acknowledges support from the U.S.
Department of Energy under the DOE/BES Program in the Applied Mathematical
Sciences, Contract KC-07-01-01 at LANL. W.\ Klein acknowledges support from
the ASC Modeling Program at LANL.
\end{acknowledgments}

\end{document}